\documentstyle[prd,aps,eqsecnum]{revtex}
 
\begin{document}
\draft
%\twocolumn[\hsize\textwidth\columnwidth\hsize\csname
%@twocolumnfalse\endcsname
%\preprint{PU-RCG-97/11,~SUSX-TH-97/13}
\title{Symmetries for generating string cosmologies
\footnote{To appear in Physical Review D.
Copyright 1998 by The American Physical Society}
}
\author{E. J. Copeland$^1$, James E. Lidsey$^{1,2,3}$ and David Wands$^4$}
\address{$^1$Centre for Theoretical Physics, University of Sussex, 
Brighton, BN1 9QH,\ U.K. \\
$^2$Astronomy Centre, 
University of Sussex, Brighton, BN1 9QH,\ U.K.  \\
$^3$Astronomy Unit, School of Mathematical Sciences,
Queen Mary and Westfield, Mile End Road, London, E1 4NS, U.K. \\
$^4$School of Computer Science and Mathematics, 
University of Portsmouth, Portsmouth, PO1 2EG,\ U.K.}
\date{\today}
\maketitle
\begin{abstract}
We discuss the symmetry properties of the low--energy effective action
of the type IIB superstring that may be employed to derive
four--dimensional solutions. A truncated effective action,
compactified on a six--torus, but including both
Neveu/Schwarz--Neveu/Schwarz and Ramond--Ramond field strengths, can
be expressed as a non-linear sigma model which is invariant under
global SL(3,R) transformations. This group contains as a sub--group
the SL(2,R) symmetry of the ten--dimensional theory and a discrete
Z$_2$ reflection symmetry which leads to a further SL(2,R)
sub--group. The symmetries are employed to analyse a general class of
spatially homogeneous cosmological solutions with non--trivial
Ramond--Ramond fields.
\end{abstract}
 
% \vskip2pc]

\section{Introduction}

String theory has entered a new era. 
There is now strong evidence that the five separate perturbative 
string theories are related non--perturbatively by duality symmetries
\cite{reviewduality,giveon,hulltownsend,hull,vafa}. 
This suggests that the five theories may 
have a common origin in a more fundamental theory (M--theory), the  
low--energy limit of which is eleven--dimensional (11--D), 
$N=1$ supergravity. 
The symmetries of the type II string theories in particular 
have a number of interesting
features~\cite{hulltownsend}. The equations of motion of
10--D type IIB supergravity are symmetric under
global SL(2,R) transformations. The subgroup SL(2,Z) is
the conjectured S--duality of the type IIB superstring
\cite{hulltownsend}.  This relates the strong-- and weak--coupling
regimes of the theory and interchanges Neveu/Schwarz--Neveu/Schwarz
(NS--NS) and Ramond--Ramond (RR) charges.  

Type IIA and IIB supergravity theories are equivalent after toroidal
compactification \cite{equiv}. Compactification of the low-energy
effective type IIB action on a six--torus results in $N=8$
supergravity, which may also be derived by toroidally compactifying
11--D, $N=1$ supergravity \cite{cj}. This contains seventy scalar and
pseudo--scalar fields that parametrize the ${\rm E}_{7(7)}/[{\rm
SU}(8)/{\rm Z}_2]$ coset and the field equations are invariant under
the global action of the group ${\rm E}_{7(7)}$.  The discrete
subgroup ${\rm E}_{7(7)}({\rm Z})$ is the conjectured U--duality of
the type II superstring and contains the T--duality group ${\rm
O}(6,6;{\rm Z})$ and S--duality group ${\rm SL(2,Z)}$
\cite{hulltownsend}.

The purpose of the present paper is to consider a truncated sector of
4--D, $N=8$ supergravity.  We derive a 4--D low--energy effective
action that includes non--trivial interaction terms between the NS--NS
and RR form fields, but maintains enough simplicity that exact
solutions to the field equations can be found analytically and
analysed in detail.  A study of the symmetries of such an effective
action is important because these provide a powerful method for
generating new string solutions from existing vacuum solutions.  
It is known that there exists a global ${\rm O}(d,d)$ symmetry in the
field equations when a string background admits $d$ Abelian isometries
\cite{mei,mahsch}.  (For an extensive review, see, e.g.,
Ref. \cite{giveon}). For example, this symmetry has been employed to
generate curved backgrounds with a time-dependent tensor potential from
flat solutions \cite{gas} and homogeneous cosmologies \cite{gas1}.  
On the other hand the presence of an SL(2,R) symmetry~\cite{NSdual} in
the NS-NS sector of the theory can been used to generate a
time-dependent field-strength~\cite{CLW94,kaloper,CEW97,CLW97}.
Such 4--D solutions derived from the low--energy effective action are
interesting because they represent a first approximation to string
backgrounds that are exact to all orders. Furthermore, the possible
role played by RR fields in cosmology may be investigated using such
an action.

In view of this, we compactify the low--energy effective type IIB
theory on an isotropic six--torus and only include the variation of
the moduli and form--fields on the external 4--D spacetime. 
In four dimensions, the three--form field strengths are dual to
one--forms and this correspondence is employed to derive a dual action
in terms of five scalar and pseudo--scalar `axion' fields. We find
that these fields parametrize an SL(3,R)/SO(3) coset.  The global
SL(2,R) symmetry of the 10--D supergravity action is preserved in 4--D
as a subgroup of the SL(3,R) symmetry.  We also identify a discrete
${\rm Z}_2$ symmetry, corresponding to a subgroup of the full O(6,6;Z)
T--duality, which leads to a further SL(2,R) symmetry that may be
viewed as a `mirror' image of the original SL(2,R) symmetry.

The paper is organised as follows. We derive the dual effective action
in Section II and discuss its global symmetries in Section III. In
Section IV, we present a general class of homogeneous cosmological
solutions as an example of how 4--D spacetimes with non--trivial
NS--NS and RR fields may be derived by employing the symmetries of the
action. Cosmological solutions obtained by employing the full SL(3,R)
symmetry~\cite{prep} generalise results previously obtained from a
single SL(2,R) transformation acting on the NS--NS
solutions~\cite{PS,Feinstein}.

\section{The effective action}

The low--energy limit of the type IIB 
superstring is 10--D, $N=2$ chiral supergravity 
\cite{IIBeffectiveaction}. The 
bosonic massless excitations 
in the NS--NS sector are 
the dilaton, $\Phi$, the metric,  $g_{MN}$, 
and the antisymmetric, two--form potential,  
$B^{(1)}_{MN}$. The RR sector contains a scalar axion field, 
$\chi$, a two--form potential,  $B^{(2)}_{MN}$, 
and a four--form potential, $D_{MNPQ}$. The equation 
of motion for the four--form cannot 
be derived from a covariant 10--D action and we 
therefore assume that this field vanishes \cite{cov}. 
The field equations for 
the remaining degrees of freedom can be derived by extremizing 
the action~\cite{hull1}: 
\begin{equation}
\label{IIB}
S_{\rm IIB} =\int d^{10} x \sqrt{-g_{10}} \left\{ e^{-\Phi} 
\left[ R_{10} +\left( \nabla \Phi \right)^2 
-\frac{1}{12} (H^{(1)})^2  \right]
% \right. \nonumber \\ \left.
 -\frac{1}{2} \left( \nabla \chi \right)^2 -\frac{1}{12} 
\left(  H^{(1)} \chi +H^{(2)} \right)^2 \right\} ,
\end{equation}
where $R_{10}$ is the Ricci curvature scalar, 
$g_{10} \equiv {\rm det} (g_{MN})$ and\footnote{In 
this paper, Upper case Latin indices
take values in the range $A=(0, 1, \ldots , 9)$, lower case Latin
indices vary from $a=(4, 5, , \ldots 9)$ and Greek indices are given
by $\mu = (0, 1, 2, 3)$. The signature of spacetime is $(-, +, +, \ldots ,
+)$.}
\begin{equation}
\label{2form}
H^{(i)}_{MNP} 
\equiv \partial_{[M} B^{(i)}_{NP]} 
\end{equation}
are the field strengths of the two--form potentials $B^{(i)}_{MN}$.
The RR terms do not couple directly to the 10--D dilaton in
the string frame, and the dilaton field is 
minimally coupled to the metric in the conformally
related,  10--D Einstein frame:
\begin{equation}
\label{conften}
\hat{g}_{MN} = e^{-\Phi/4} g_{MN}. 
\end{equation}

We then have 
\begin{equation}
\label{IIBEinstein}
S_{\rm IIB} =\int d^{10} x \sqrt{-\hat{g}_{10}} \left\{   
 \hat{R}_{10} - {1\over8} \left( \hat\nabla \Phi \right)^2
 -\frac{1}{2} e^{\Phi} \left( \hat\nabla \chi \right)^2 
%\right. \nonumber \\ \left. 
 -\frac{1}{12} e^{\Phi/2} ( \hat{H}^{(1)} )^2 
-\frac{1}{12} e^{-\Phi/2}
 \left( \hat{H}^{(1)} \chi + \hat{H}^{(2)} \right)^2 \right\}
\end{equation}
and 
Eq. (\ref{IIBEinstein}) can be written
as~\cite{hulltownsend,2bsym} 
\begin{equation}
\label{10DSLaction}
S_{\rm IIB} =\int d^{10} x \sqrt{-\hat{g}_{10}} \left\{  
 \hat{R}_{10}
 + {1\over4} {\rm Tr} \left( \hat\nabla M \hat\nabla M^{-1} \right)
 - \frac{1}{12} \hat{H}^T M \hat{H} \right\}  ,
\end{equation}
where
\begin{equation}
\label{matrixM}
M \equiv   \left( \begin{array}{cc} e^{\Phi/2} &  \chi e^{\Phi/2} \\ 
\chi e^{\Phi/2} & e^{-\Phi/2} + \chi^2 e^{\Phi/2} \end{array} \right) ,
\end{equation}
and
\begin{equation}
\label{vectorH}
\hat{H}  = \left( \begin{array}{c} 
\hat{H}^{(2)} \\ \hat{H}^{(1)} \end{array} \right) .
\end{equation}

It follows that $M^TJM=J $, where 
\begin{equation}
J  \equiv   \left( \begin{array}{cc} 0 & 1 \\ 
-1 & 0 \end{array} \right) , \qquad J^2 =-I
\end{equation}
is the SL(2,R) metric. Thus,  $M$ is a member of the group 
SL(2,R) and the action (\ref{10DSLaction}) is invariant under the global 
SL(2,R) transformation \cite{hulltownsend,2bsym}: 
\begin{equation}
\label{schwarz}
\bar{M} = \Sigma M \Sigma^T, \qquad \bar{\hat{g}}_{MN} = \hat{g}_{MN} , 
\qquad \bar{\hat{H}} = \left( \Sigma^T \right)^{-1} \hat{H}  ,
\end{equation}
where 
\begin{equation}
\label{sigma}
\Sigma \equiv \left( \begin{array}{cc} D & C \\ 
B  & A \end{array} \right) , \qquad AD-BC =1 .
\end{equation}
The invariance of the 10--D Einstein metric $\hat{g}_{MN}$ implies 
that the string metric $g_{MN}$ transforms under Eq. (\ref{schwarz}) as
\begin{equation}
\bar{g}_{MN} e^{-\bar{\Phi} /4} = g_{MN} e^{-\Phi /4}  .
\end{equation}
The four--form transforms as a singlet. 

\subsection{Reduced 4--D action}

The toroidal compactification of the type IIB theory (\ref{IIB}) has
recently been discussed \cite{hull1,mah,roy,bbo,a}. 
Maharana \cite{mah} and Roy 
\cite{roy} compactified down to $D$ dimensions 
and showed how the SL(2,R) symmetry discussed above is
respected in lower dimensions. We compactify to four dimensions 
with the simplest toroidal {\em ansatz}:
\begin{equation}
\label{ansatz}
ds_{10}^2 =g_{\mu\nu} (x) dx^{\mu} dx^{\nu} 
+e^{y(x)/\sqrt{3}} \delta_{ab} dX^a dX^b   ,
\end{equation}
where $y$ is a modulus field describing the volume of the internal
space.  This field represents the `breathing mode'
of the internal space. We assume that all fields are independent of the
internal coordinates $X^a$. 
We do not consider the vector fields arising
from the metric components or the moduli originating from the
compactification of the three--forms. Although this
is a more severe truncation of the toroidally compactified action than 
the one considered by Maharana~\cite{mah} and
Roy~\cite{roy}, it has the advantage that the interactions between the
NS--NS and RR fields can be quantitatively analysed. Each form field
is reduced to a single degree of freedom and this allows
valuable insight to be gained into the role played by the RR sector.

The reduced 4--D  effective action in the string frame is given by
\begin{eqnarray}
\label{4daction}
S_4&=& \int d^4 x \sqrt{-g} \left\{ e^{-\varphi} \left[ R + 
\left( \nabla\varphi \right)^2
 - {1\over2} \left( \nabla y \right)^2 
-\frac{1}{12} (H^{(1)})^2 \right]
 \right. \nonumber \\ 
&& \left. \qquad 
-\frac{1}{2} e^{\sqrt{3}y} \left( \nabla \chi \right)^2 -\frac{1}{12} 
e^{\sqrt{3}y} (H^{(1)} \chi +H^{(2)}  )^2 \right\}   ,
\end{eqnarray}
where
\begin{equation}
\label{shifteddilaton}
\varphi \equiv \Phi - \sqrt{3}y
\end{equation}
represents the effective dilaton in the 4--D spacetime. The corresponding 
action in the 4--D Einstein frame
\begin{equation}
\label{conformal}
\tilde{g}_{\mu\nu} = e^{-\varphi} g_{\mu\nu} 
\end{equation}
is then given by 
\begin{eqnarray}
\label{confaction}
S_4=\int d^4 x \sqrt{-\tilde{g}} \left\{ \tilde{R}_4 -\frac{1}{2} 
\left( \tilde{\nabla} \varphi \right)^2 - {1\over2} \left( \tilde{\nabla} 
y \right)^2 -\frac{1}{12} e^{-2\varphi} ( \tilde{H}^{(1)} )^2
\right. \nonumber \\
\left. 
- \frac{1}{2} e^{\sqrt{3}y +\varphi} \left( 
\tilde{\nabla} \chi \right)^2 
-\frac{1}{12} e^{\sqrt{3}y -\varphi} ( \tilde{H}^{(1)} \chi + 
\tilde{H}^{(2)} )^2
\right\}   .
\end{eqnarray}
Note that the 4--D Einstein frame metric, Eq. (\ref{conformal}),  
differs from  the corresponding part 
of the 10--D Einstein frame metric  (\ref{conften})
by the conformal factor
\begin{equation}
\label{10Dfactor}
\Omega^2=e^{\Phi/4-\varphi}=e^{\sqrt{3}(y-\sqrt{3}\varphi)/4}  .
\end{equation}

\subsection{The Dual Action}

In what follows  we shall always refer to the metric 
in the 4--D Einstein
frame and so we drop tildes for notational simplicity.
The field equations for the three--forms in the 4--D Einstein frame are
given by
\begin{eqnarray}
\label{hfield2}
\nabla_{\mu} \left[ e^{\sqrt{3}y -\varphi} \left( \chi H^{(1) \, 
\mu\nu\lambda} +
H^{(2) \, \mu\nu\lambda} \right) \right] =0 \\
\label{hfield1}
\nabla_{\mu} \left[ e^{-2\varphi} H^{(1) \, \mu\nu\lambda} + \chi 
e^{\sqrt{3}y -\varphi} \left( \chi H^{(1) \, \mu\nu\lambda} +
H^{(2) \, \mu\nu\lambda}
\right) \right]  = 0 .
\end{eqnarray}
In four dimensions the three--form field strengths are dual to one--forms:
\begin{eqnarray}
\label{dualK}
H^{(1)}_{\mu\nu\lambda} \equiv \epsilon_{\mu\nu\lambda\kappa} K^{\kappa}  \\
\label{dualJ}
H^{(2)}_{\mu\nu\lambda} \equiv \epsilon_{\mu\nu\lambda\kappa} J^{\kappa}  ,
\end{eqnarray}
where $\epsilon_{\mu\nu\lambda\kappa}$ is the covariantly constant
four--form. The field equations (\ref{hfield2}) and 
(\ref{hfield1}) take the form 
\begin{eqnarray}
\label{Zeom}
\epsilon^{\mu\nu\lambda\kappa} \nabla_\mu 
 \left[ e^{\sqrt{3}y -\varphi} \left( \chi K_\kappa +
  J_\kappa \right) \right] =0 \\
\label{Yeom}
\epsilon^{\mu\nu\lambda\kappa} \nabla_\mu
 \left[ e^{-2\varphi} K_\kappa + \chi 
 e^{\sqrt{3}y -\varphi} \left( \chi K_\kappa + J_\kappa \right) \right]  = 0 
\end{eqnarray}
when written in terms of the dual one--forms. 

These dual forms may be written in terms
of the gradients of two pseudo--scalar `axion' fields.
Equation~(\ref{Zeom}) requires that
\begin{equation}
\label{Jeq}
e^{\sqrt{3}y -\varphi} \left( \chi K_\kappa + J_\kappa \right)
 = \nabla_\kappa \sigma_2   ,
\end{equation}
where $\sigma_2$ is any scalar function. Substituting this into
Eq.~(\ref{Yeom}) implies that 
\begin{equation}
\label{Keq}
e^{-2\varphi} K_\kappa + \chi \nabla_\kappa\sigma_2
 = \nabla_\kappa \sigma_1  ,
\end{equation}
where $\sigma_1$ 
is a second arbitrary scalar 
function. These together imply that the field equations 
(\ref{hfield2}) and (\ref{hfield1}) are automatically
satisfied by
\begin{eqnarray}
\label{defH1}
H^{(1) \, \mu\nu\lambda} &=& \epsilon^{\mu\nu\lambda\kappa}
 e^{2\varphi} \left( \nabla_\kappa\sigma_1 - \chi\nabla_\kappa\sigma_2
 \right) \\
\label{defH2}
H^{(2) \, \mu\nu\lambda} &=& \epsilon^{\mu\nu\lambda\kappa}
  \left[ e^{\varphi-\sqrt{3}y}\nabla_\kappa\sigma_2 - \chi e^{2\varphi}
 \left( \nabla_\kappa\sigma_1 - \chi\nabla_\kappa\sigma_2 \right)
 \right]  .
\end{eqnarray}

It should be emphasised that the definitions of both scalar fields
$\sigma_i$ are arbitrary up to a redefinition $\sigma_i \rightarrow
\sigma_i +f_i$, where $f_i$ represents an arbitrary scalar function.
In this sense, therefore, there is no unique definition of the
pseudo--scalar axion fields. However, when $H^{(2)}_{\mu\nu\lambda} =
\chi=0$, we see that our definition of $\sigma_1$ coincides with the
usual definition of the axion that is dual to the NS--NS three--form
field strength~\cite{NSdual,CLW94,CLW95,CEW97}.

Although the field
equations~(\ref{hfield2}) and~(\ref{hfield1})
for the three--forms are now automatically satisfied by 
the dual ansatz, we must also impose the Bianchi identities 
\begin{equation}
\label{bianchiidentity}
\nabla_{[\mu} H^{(i)}_{\nu\lambda\kappa]} \equiv 0 
\end{equation}
that arise because the three--form field strengths are defined 
in terms of the  
gradients of two--form potentials. These correspond to the 
constraint equations
\begin{eqnarray}
\label{fieldsigma1}
\nabla_{\rho} \left[ e^{2\varphi} \left( 
\nabla^{\rho} \sigma_1 -\chi \nabla^{\rho} \sigma_2 \right) 
\right] =0 \\
\label{fieldsigma2}
\nabla_{\rho} \left[ e^{-\sqrt{3}y +\varphi} \nabla^{\rho} \sigma_2
-\chi e^{2\varphi} \left( \nabla^{\rho} \sigma_1 -\chi \nabla^{\rho} 
 \sigma_2 \right) \right] =0
\end{eqnarray}
on the fields $\sigma_i$. They are interpreted in the dual ansatz as field 
equations that can be derived from
the effective action 
\begin{eqnarray}
\label{solitonicaction}
S_{4*}&=&\int d^4 x \sqrt{-g} \left[ R -\frac{1}{2} 
\left( \nabla \varphi \right)^2 - {1\over2} \left( 
\nabla y \right)^2  -\frac{1}{2} e^{\sqrt{3}y +\varphi}
\left( \nabla \chi \right)^2 
 \right. \nonumber \\ 
&& \left. \qquad
-\frac{1}{2} e^{-\sqrt{3}y +\varphi}
\left( \nabla \sigma_2 \right)^2  -\frac{1}{2} 
e^{2\varphi} \left( \nabla \sigma_1 -\chi \nabla \sigma_2 \right)^2 
\right]   .
\end{eqnarray}

This action is not identical to the original action (\ref{confaction})
because the roles of the Bianchi identities and the field equations are
interchanged. Nevertheless,  the two descriptions
are {\em dynamically equivalent} as
long as the field equations are satisfied, and it should be emphasised
that either form of the action should only be viewed as an effective
action which reproduces the correct equations of motion. 
With this in mind, we refer to this dual action when investigating 
the symmetries of the 4--D theory. 
The equations of motion for 
the five scalar fields are then given by Eqs.~(\ref{fieldsigma1})
and~(\ref{fieldsigma2}), together with 
\begin{eqnarray}
\label{fieldphi}
\Box \varphi = {1\over2} e^{\sqrt{3}y+\varphi} (\nabla\chi)^2
 + {1\over2} e^{-\sqrt{3}y+\varphi} (\nabla\sigma_2)^2
 + e^{2\varphi} \left( \nabla\sigma_1 - \chi\nabla\sigma_2 \right)^2
\\
\label{fieldy}
\Box y  =  {\sqrt{3}\over2} e^{\sqrt{3}y+\varphi}(\nabla\chi)^2
 -  {\sqrt{3}\over2} e^{-\sqrt{3}y+\varphi}(\nabla\sigma_2)^2
\\
\label{fieldchi}
\nabla^\mu ( e^{\sqrt{3}y+\varphi}\nabla_\mu\chi )
  =  - e^{2\varphi} \nabla^\mu\sigma_2 \left( \nabla_\mu\sigma_1 -
 \chi\nabla_\mu\sigma_2 \right)   .
\end{eqnarray}

\subsection{Conserved currents}

The one--forms $K$ and $J$ defined in
Eqs.~(\ref{dualK}) and~(\ref{dualJ}) can be written in terms of the
pseudo--scalar axion fields using Eqs.~(\ref{Jeq}) and~(\ref{Keq}). This 
yields 
\begin{eqnarray}
\label{Kcurrent}
K_\mu = e^{2\varphi} ( \nabla_\mu\sigma_1 - \chi \nabla_\mu\sigma_2 )
\\
\label{Jcurrent}
J_\mu = e^{-\sqrt{3}y+\varphi} \nabla_\mu\sigma_2 - \chi K_\mu
\end{eqnarray}
and the Bianchi identities~(\ref{fieldsigma1}) and~(\ref{fieldsigma2})
simply correspond to the requirement that these currents are conserved:
\begin{equation}
\nabla^\mu K_\mu =0 \ , \qquad \nabla^\mu J_\mu =0 .
\end{equation}
In terms of the original three--form field strengths, these currents are
topologically conserved due to the Bianchi identities, but in the dual
formulation they are Noether currents conserved due to the symmetry of
the action~(\ref{solitonicaction}).

The conserved currents allow us to integrate out the kinetic terms
for the pseudo--scalar axion fields $\sigma_i$. 
This reduces the field equations~(\ref{fieldphi}--\ref{fieldchi}) to
\begin{eqnarray}
\label{redfieldphi}
\Box \varphi = {1\over2} e^{\sqrt{3}y+\varphi} (\nabla\chi)^2
 - {\partial V \over \partial\varphi}
\\
\label{redfieldy}
\Box y  =  {\sqrt{3}\over2} e^{\sqrt{3}y+\varphi}(\nabla\chi)^2
 - {\partial V \over \partial y}
\\
\label{redfieldchi}
\nabla^\mu ( e^{\sqrt{3}y+\varphi}\nabla_\mu\chi )
  = - {\partial V \over \partial\chi} \ ,
\end{eqnarray}
where the effective interaction potential, $V$, for the fields
$\varphi$, $y$ and $\chi$ is given by
\begin{equation}
\label{effV}
V = - \frac{1}{2} g^{\mu\nu}
 \left[ e^{\sqrt{3}y -\varphi} \left( J_\mu + \chi K_\mu \right) 
    \left( J_\nu + \chi K_\nu \right) 
  + e^{-2\varphi} K_\mu K_\nu \right] .
\end{equation}
This arises due to the three--form field strengths and may be expressed 
in a  SL(2,R) invariant form, as will be shown in the following section. 

Note that the field equation for the field $\chi$,
Eq.~(\ref{fieldchi}), can also be written as
\begin{equation}
\nabla^\mu ( e^{\sqrt{3}y+\varphi}\nabla_\mu\chi )
  =  - K_\mu \nabla^\mu\sigma_2   .
\end{equation}
Since $\nabla^\mu K_\mu=0$, we deduce that
\begin{equation}
\label{Lcurrent}
L_\mu = e^{\sqrt{3}y+\varphi}\nabla_\mu\chi + \sigma_2 K_\mu   ,
\end{equation}
where $\nabla^\mu L_\mu=0$. Thus, $L_\mu$ is the third conserved
current for the form--fields, independent of $K_\mu$ and $J_\mu$.
This does not allow us to simplify the
equations~(\ref{redfieldphi}--{\ref{redfieldchi}) any further as we
simply swap our ignorance of $\nabla\chi$ for our ignorance of
$\sigma_2$.  However, it is indicative of a further symmetry of the
dual action which we discuss in the following section.

\section{Symmetries of the dual action}

\subsection{SL(3,R) symmetry}

The effective action (\ref{solitonicaction}) can be expressed as 
a 4--D, non--linear sigma--model: 
\begin{equation}
\label{nonlinsigma}
S_{4*} = \int d^4 x \sqrt{-g} \left[ R -{\cal{G}}_{AB} 
(\varphi ) \nabla \varphi^A \nabla \varphi^B \right]   ,
\end{equation}
where the scalar fields $\varphi^A = (\varphi ,y ,\chi ,
\sigma_1 ,\sigma_2 )$, $(A,B)=(1,2, \dots ,  5)$
may be viewed as coordinates on the target space with metric
\begin{equation}
\label{tsmetric}
{\cal{G}} ={\cal{G}}_{AB} d\varphi^A d \varphi^B =
\frac{1}{2} d\varphi^2 + \frac{1}{2} dy^2 +\frac{1}{2} 
e^{2\varphi} \left( d\sigma_1 -\chi d\sigma_2 \right)^2 +
\frac{1}{2} e^{\varphi} \left[ e^{\sqrt{3} y} d\chi^2 
+e^{-\sqrt{3} y} d\sigma_2^2 \right]  .
\end{equation}

Eq.~(\ref{tsmetric}) is formally identical to the target space
considered by Gal'tsov, Garcia and Kechkin within the context of 5--D,
Kaluza--Klein theory admitting two commuting Killing vectors
\cite{gal}. Maison first showed that this target space represents the
SL(3,R)/SO(3) coset corresponding to a homogeneous symmetric
Riemannian space, where the group SL(3,R) acts transitively
\cite{mai}. It can be shown by employing the Gauss decomposition of
the general SL(3,R) matrix that the action (\ref{solitonicaction}) may
be written in the form~\cite{gal}
\begin{equation}
\label{Uaction}
S_{4*} =\int d^4 x \sqrt{-g} \left[ 
R +\frac{1}{4} {\rm Tr} \left[ \nabla U \nabla U^{-1} 
\right] \right] ,
\end{equation}
where 
\begin{equation}
\label{S}
U \equiv  e^{\varphi+y/\sqrt{3}} 
\left( \begin{array}{ccc} 
1 & \chi & \sigma_1 -\chi \sigma_2  \\ 
\chi & \chi^2 +e^{-\varphi-\sqrt{3}y} & \chi ( \sigma_1 -\chi \sigma_2 )
- \sigma_2 e^{-\varphi-\sqrt{3}y} \\
\sigma_1 -\chi \sigma_2 & \chi ( \sigma_1 -\chi \sigma_2 )
- \sigma_2 e^{-\varphi-\sqrt{3}y} & ( \sigma_1 -\chi \sigma_2 )^2
 + \sigma_2^2 e^{-\varphi-\sqrt{3}y} + e^{-2\varphi}
\end{array} \right)
\end{equation}
is a symmetric SL(3,R) matrix.

We may conclude, therefore, that the dual action
(\ref{solitonicaction}) is invariant under global SL(3,R)
transformations. We now consider the relevant SL(2,R) subgroups that
prove useful in generating non--trivial solutions with RR and NS--NS
fields.

\subsection{SL(2,R)$_\chi$ symmetry}

The effective 4--D action (\ref{4daction}) still exhibits the global
SL(2,~R) symmetry of the full 10--D action manifest in
Eq.~(\ref{10DSLaction}) ~\cite{mah,roy}. This becomes apparent by defining  
new scalar fields: 
\begin{eqnarray}
\label{ufield}
{1\over2}\Phi \equiv u \equiv \frac{1}{2} \varphi + {\sqrt{3}\over2} y \\
\label{vfield}
v \equiv \frac{\sqrt{3}}{2} \varphi - {1\over2} y  .
\end{eqnarray}
The action given in Eq.~(\ref{solitonicaction}) then takes the form
\begin{eqnarray}
\label{uaction}
S_{4*}&=&\int d^4 x \sqrt{-g} \left[ R - \frac{1}{2} \left( \nabla u \right)^2 
-\frac{1}{2} e^{2u} \left( \nabla \chi \right)^2 - 
\frac{1}{2} \left( \nabla v \right)^2
 \right. \nonumber \\ 
&& \left. \qquad
 - \frac{1}{2} 
e^{\sqrt{3} v} \left( e^{-u} (\nabla\sigma_2)^2 + e^{u} ( \chi\nabla\sigma_2 -
\nabla\sigma_1 )^2 
\right) \right]   .
\end{eqnarray}

The SL(3,R) matrix $U$ given in Eq.~(\ref{S}) can be written as
\begin{equation}
U = \left( 
\begin{array}{cc}
e^{v/\sqrt{3}}M & -e^{v/\sqrt{3}} M\sigma \\
-e^{v/\sqrt{3}} \sigma^T M & e^{-2v/\sqrt{3}} + e^{v/\sqrt{3}}
\sigma^TM\sigma 
\end{array}
\right) \ ,
\end{equation}
where the symmetric $2 \times 2$ matrix $M$ is given in
Eq.~(\ref{matrixM}), and we have defined the vector
\begin{equation}  
\label{sigmavector}
\sigma \equiv \left( \begin{array}{c} -\sigma_1 \\ 
  \sigma_2 \end{array} \right) \ .
\end{equation}
This implies that Eq. (\ref{Uaction}) may be written as
\begin{equation}
\label{dualSLchiaction}
S_{4*}=\int d^4 x \sqrt{-g} \left[ R + \frac{1}{4} {\rm Tr} 
\left[ \nabla M \nabla M^{-1} \right] -\frac{1}{2} \left( \nabla 
v \right)^2
 -\frac{1}{2} e^{\sqrt{3} v} \nabla\sigma^T M \nabla\sigma \right]  .
\end{equation}

The action remains invariant under the sub-group
\begin{equation}
\bar{U} = \tilde\Sigma_\chi U \tilde\Sigma_\chi \ ,
\end{equation}
where
\begin{equation}
\tilde\Sigma_\chi = \left(
\begin{array}{cc}
\Sigma & 0 \\
0 & 1
\end{array}
\right) \ ,
\end{equation}
and this corresponds to the SL(2,R) transformation
\begin{equation}
\label{sl2rchi}
\bar{M} = \Sigma M \Sigma^T, \qquad \bar{g}_{\mu\nu} = 
g_{\mu\nu} , \qquad \bar{\sigma} = \left( \Sigma^T \right)^{-1} \sigma, 
\qquad \bar{v} =v .
\end{equation}

The transformation (\ref{sl2rchi}) acts non--linearly on the scalar fields
$u$ and $\chi$: 
\begin{eqnarray}
\label{newu}
e^{\bar{u}} = C^2 e^{-u} + (D +C \chi )^2 e^u \\
\label{newchi}
\bar{\chi} e^{\bar{u}} = AC e^{-u} + (B + A \chi )(D + C \chi ) e^u 
\end{eqnarray}
and the pseudo--scalar axion fields transform as
\begin{eqnarray}
\bar\sigma_1 = A\sigma_1 + B\sigma_2 \\
\label{endsl2rchi}
\bar\sigma_2 = C\sigma_1 + D\sigma_2  .
\end{eqnarray}

When $C=-B =-1$ and $A=D=0$, Eq. (\ref{sl2rchi}) interchanges the two
axion fields, $\sigma_i$, and inverts the 10--D string coupling
$\bar{g}_s =\exp( {\bar{\Phi}} ) = g_s^{-1} = \exp ( -\Phi )$ that is
defined in terms of the 10--D dilaton, $\Phi=2u$.  Thus, the
strongly--coupled regime of the theory is mapped onto the
weakly--coupled one, and vice--versa. The
effective 4--D dilaton field (\ref{shifteddilaton})
transforms as $\bar{\varphi}=-\varphi/2+\sqrt{3}v/4$. 

We refer to this as the SL(2,R)$_\chi$ symmetry. It is the SL(2,R)
symmetry of the 10--D theory written in terms of the 4--D
variables~\cite{mah,roy}. 
The field $v$ determines the conformal
factor (\ref{10Dfactor}) that relates the 4--D Einstein metric to the
corresponding part of the 10--D Einstein metric (\ref{conften}). The
invariance of both $v$ and the 4--D metric $g_{\mu\nu}$ implies that
the corresponding components,
$\hat{g}_{\mu\nu}=e^{\sqrt{3}v/2}g_{\mu\nu}$, of the 10--D Einstein
metric are also invariant. The radius of the internal
space in the 10--D Einstein frame is 
$e^{-v/\sqrt{3}}$ and the complete 10--D Einstein metric
(\ref{conften}) is therefore invariant, as in Eq.~(\ref{schwarz}).

The effective interaction (\ref{effV}) may also be written in a SL(2,R) 
invariant form:
\begin{equation}
\label{SLchipot}
V(u,v,\chi)
 = - {1\over2} g^{\mu\nu} e^{-\sqrt{3}v} {\cal J}_\mu^T M {\cal J}_\nu    ,
\end{equation}
where 
\begin{equation}
{\cal J}_\mu = \left( \begin{array}{c} J_\mu \\ 
  K_\mu \end{array} \right) 
\end{equation}
transforms as $\bar{{\cal J}}_\mu = \left( \Sigma^T \right)^{-1} {\cal
J}_\mu$, and thus the covariantly conserved currents transform as
\begin{eqnarray} 
\bar{J}_\mu &=& A J_\mu - B K_\mu \\
\bar{K}_\mu &=& -C J_\mu + D K_\mu  .
\end{eqnarray}
This provides an interesting example of a non--trivial $(V\neq {\rm
constant})$ SL(2,R) invariant interaction for the dilaton, despite the
fact that there is no SL(2,R) invariant potential that can be derived
from the dilaton alone~\cite{mah}. The three--form field strengths can
provide an effective interaction potential for the 10--D dilaton,
$\Phi=2u$, because they too transform under the SL(2,R)
transformation.

The form of this interaction suggests that it may be possible to
stabilise the 10--D dilaton due to interactions with the axion fields,
as was recently noted in a similar context by Lukas, Ovrut and
Waldram~\cite{stable}.  However, this is only possible if some other
mechanism (in addition to the fields considered in this paper)
operates to stabilise $v$ (the size of the compact space in the 10--D
Einstein frame) and prevents the prefactor $e^{-\sqrt{3}v}\to0$ in
Eq.~(\ref{SLchipot}).

Finally, we note that although a general SL(2,R) matrix of the form
given in Eq.~(\ref{sigma}) has three independent real parameters,
there is a two-dimensional sub-group,
\begin{equation}
\label{sigma2d}
\Sigma_0 \equiv \left( \begin{array}{cc} A^{-1} & 0 \\ 
B  & A \end{array} \right) ,
\end{equation}
which leaves the Lagrangian (\ref{solitonicaction}) invariant term by
term. These transformations correspond either to a constant shift or
rescaling of the axion fields, such that the three
four-vectors, $e^{-\varphi}K_\mu$, 
$e^{(-\varphi+\sqrt{3}y)/2}(J_\mu+\chi K_\mu)$ and
$e^{(-\varphi-\sqrt{3}y)/2}(L_\mu-\sigma_2K_\mu)$ remain invariant.
Thus the only non-trivial transformation is the ``boost''
\begin{equation}
\label{sigma1d}
\Sigma_1 \equiv \left( \begin{array}{cc} \cosh\theta & \sinh\theta \\ 
\sinh\theta  & \cosh\theta \end{array} \right) ,
\end{equation}
which introduces at most one new parameter $\theta$.

We can use the SL(2,R) symmetry of the action to generate new
four-dimensional solutions of the field equations.  As an example in
Section~IV we will describe the homogeneous cosmological solutions
with non-trivial $\chi$ field by applying the transformation given in
Eq.~(\ref{sl2rchi}) to the homogeneous dilaton-moduli-vacuum
solutions.

\subsection{Z$_2$ and SL(2,R)$_{\sigma_2}$ symmetry}

The importance of the dual action is that a further SL(2,R) symmetry
may be uncovered. The NS--NS sector of the reduced
action~(\ref{4daction}) is invariant under the `T--duality'
transformation $\bar{y}=-y$, corresponding to an inversion of the
internal space. This ${\rm Z}_2$ symmetry can be extended to the RR
sector of the theory and the dual action~(\ref{solitonicaction}) is
symmetric under the discrete transformation
\begin{equation} 
\label{td}
\bar{y} =-y , \qquad \bar{\chi} = \sigma_2 , \qquad 
\bar{\sigma}_2 = \chi , \qquad \bar{\sigma}_1 =-\sigma_1+\chi
\sigma_2 ,
\end{equation}
where the 4--D dilaton, $\varphi$, and 4--D Einstein frame metric
remain invariant. Note, however, that because the volume of the internal
space changes, the 10--D Einstein frame metric 
(related to the 4--D Einstein
frame metric by the conformal factor given in Eq.~(\ref{10Dfactor})) is {\em
not} invariant under this transformation.
In terms of the conserved axion currents, defined in Eqs.~(\ref{Jcurrent}),
(\ref{Kcurrent}) and~(\ref{Lcurrent}), the
reflection symmetry (\ref{td}) corresponds to
\begin{equation}
\bar{y} =-y , \qquad \bar{K}_\mu = -K_\mu, \qquad 
\bar{J}_\mu = L_\mu , \qquad \bar{L}_\mu = J_\mu .
\end{equation}

This reflection symmetry implies the existence of an
alternative SL(2,R) symmetry in the dual action which can be obtained
from a combination of the SL(2,R)$_\chi$ transformation given in
Eq.~(\ref{sl2rchi}) plus the reflection symmetry in Eq.~(\ref{td}).
Analogously to Eqs.~(\ref{ufield}) and~(\ref{vfield}), but with
$y\to-y$, we introduce the new scalar fields:
\begin{eqnarray}
\label{wfield}
w \equiv \frac{1}{2} \varphi - {\sqrt{3}\over2} y \\
\label{xfield}
x \equiv \frac{\sqrt{3}}{2} \varphi + {1\over2} y  .
\end{eqnarray}
The dual effective action, 
Eq.~(\ref{solitonicaction}), then takes the form 
\begin{eqnarray}
\label{waction}
S_{4*} &=& 
 \int d^4 x \sqrt{-g} \left[ R - \frac{1}{2} \left( \nabla w \right)^2 
-\frac{1}{2} e^{2w} \left( \nabla \sigma_2 \right)^2 - 
\frac{1}{2} \left( \nabla x \right)^2
 \right. \nonumber \\
&& \left. \qquad
 -\frac{1}{2} e^{\sqrt{3}x} \left( 
e^{w} \left( \nabla \sigma_1 -\chi \nabla \sigma_2 \right)^2 
+e^{-w} \left( \nabla \chi \right)^2 \right) \right]  .
\end{eqnarray}
Defining the symmetric $2 \times 2$ matrix: 
\begin{equation}
\label{Nmatrix}
P \equiv   \left( \begin{array}{cc} e^{w} &  \sigma_2 
e^{w} \\ 
\sigma_2  e^{w} & e^{-w} +
\sigma_2^2 e^{w} \end{array} \right)
\end{equation}
and the vector:
\begin{equation}
\tau \equiv \left( \begin{array}{c} \sigma_1-\chi\sigma_2 \\ 
   \chi \end{array} \right) 
\end{equation}
allows us to express this action as 
\begin{equation}
\label{wacions}
S_{4*} = \int d^4 x \sqrt{-g} \left[ R +\frac{1}{4} {\rm Tr} 
\left[ \nabla P \nabla P^{-1} \right] -\frac{1}{2} \left( \nabla 
x \right)^2
 -\frac{1}{2} e^{\sqrt{3} x} \nabla\tau^T P \nabla\tau \right]  .
\end{equation}
This is manifestly invariant under the  SL(2,R) transformation
\begin{equation}
\label{sl2rsigma2}
\bar{P} = \Sigma' P \Sigma'^T, \qquad \bar{g}_{\mu\nu} = 
g_{\mu\nu} , \qquad \bar\tau = \left( \Sigma'^T \right)^{-1} \tau
\qquad \bar{x} = x 
\end{equation}
and this implies that
\begin{eqnarray}
\label{neww}
e^{\bar{w}} &=& C'^2 e^{-w} + (D' +C' \sigma_2 )^2 e^{w} \\
\label{newsigma2}
\bar{\sigma_2} e^{\bar{w}} &=& (B' + A' \sigma_2 )(D' + C' \sigma_2 ) e^{w} 
+A'C' e^{-w} \\ 
\label{chisigma2}
\bar\chi &=& -C'(\sigma_1-\sigma_2\chi) + D'\chi\\
\label{sigma1sigma2}
\bar\sigma_1-\bar\chi\bar\sigma_2 &=& A'(\sigma_1-\sigma_2\chi) - B'\chi  .
\end{eqnarray}

We refer to this as the  SL(2,R)$_{\sigma_2}$ symmetry of the action. 
It should be emphasised that this is not the 10--D ${\rm
SL(2,R)}_{\chi}$ symmetry  recast in terms of the 4--D action.  The
${\rm SL(2,R)}_{\sigma_2}$ transformation mixes the $\sigma_2$ axion
field with $w$. This latter field is not the 10--D dilaton,
because it includes an 
additional  contribution from the modulus field $y$. Thus, 
the radius of the internal dimensions transforms non-trivially and the
10--D Einstein metric is not invariant under~(\ref{sl2rsigma2}). 

Comparison of Eq.~(\ref{ufield}) with Eq.~(\ref{wfield}) and
Eq.~(\ref{vfield}) with Eq.~(\ref{xfield}) implies that the discrete
transformation $y \leftrightarrow -y$ is equivalent to $u
\leftrightarrow w$ and $v \leftrightarrow x$. Moreover,
Eqs.~(\ref{matrixM}) and~(\ref{Nmatrix}) imply that the
reflection symmetry (\ref{td}) is equivalent to $M \leftrightarrow
P$. Thus, the ${\rm SL(2,R)}_{\sigma_2}$
symmetry transformation is formally equivalent to the ${\rm Z}_2$
transformation (\ref{td}), followed by the ${\rm SL(2,R)}_{\chi}$
transformation (\ref{sl2rchi}), followed by another ${\rm Z}_2$
transformation (\ref{td}).

Finally, we recall that we were able to employ the conserved currents
for $\sigma_i$ to provide reduced equations of motion for $\varphi$,
$y$ and $\chi$ in Eqs.~(\ref{redfieldphi}--\ref{redfieldchi}), where
the effective interaction potential~Eq.~(\ref{SLchipot}) is invariant
under the SL(2,R)$_\chi$ transformation. The reflection
symmetry~(\ref{td}) implies that analogous reduced field equations
for $\varphi$, $y$ and $\sigma_2$ may be derived by
employing the conserved currents for $\sigma_1$ and
$\chi$. Substituting the definitions of $K_\mu$ and $L_\mu$,
Eqs.~(\ref{Kcurrent}) and~(\ref{Lcurrent}), into the equations of
motion implies that
\begin{eqnarray}
\label{redfieldw}
\Box w &=& e^{2w} (\nabla\sigma_2)^2 - {\partial W \over \partial\varphi}
\\
\label{redfieldx}
\Box x  &=& - {\partial W \over \partial y}
\\
\label{redfieldsg2}
\nabla^\mu ( e^{2w}\nabla_\mu\sigma_2 )
  &=& - {\partial W \over \partial\chi} \ ,
\end{eqnarray}
where the effective interaction potential $W$ between the fields
$\varphi$, $y$ and $\sigma_2$ is given by
\begin{equation}
W(w,x,\sigma_2) = -\frac{1}{2} e^{\sqrt{3} x} \nabla\tau^T P
\nabla\tau
 = -\frac{1}{2} g^{\mu\nu} e^{-\sqrt{3}x} {\cal L}_\mu^T P {\cal L}_\nu
\end{equation}
and
\begin{equation}
\label{ldef}
{\cal L}_\mu \equiv  \left( \begin{array}{c} L_\mu \\ 
  -K_\mu \end{array} \right) \ .
\end{equation}
${\cal L}_\mu$ transforms as $\bar{{\cal
L}}_\mu=(\Sigma'^T)^{-1}L_\mu$ under the SL(2,R)$_{\sigma_2}$
transformation in Eq.~(\ref{sl2rsigma2}) and the conserved currents
therefore transform as
\begin{eqnarray}
\bar{L}_\mu &=& A' L_\mu + B' K_\mu \\
\bar{K}_\mu &=& C' L_\mu + D' K_\mu   .
\end{eqnarray}

\section{Four--dimensional homogeneous cosmologies}

The symmetry transformations described in Section~III
may be employed to generate solutions with non--trivial 
NS--NS and RR form fields from {\em any} given solution 
to the dilaton--moduli--vacuum field equations. 
4--D cosmological solutions are of particular interest as they might
describe the early evolution of our own universe. 
Homogeneous and isotropic Friedmann--Robertson--Walker (FRW) 
solutions were presented in~\cite{prep}.
In this Section we extend the analysis to
spatially homogeneous but anisotropic Bianchi universes. These 
models admit a three--dimensional Lie group of isometries, 
$G_3$, acting on the surfaces of homogeneity. The space--like 
surfaces have a metric, $h_{ab}=h_{ab}(t)$ $(a,b = 1,2,3)$, where $t$ 
represents cosmic time and the four--dimensional line 
element in the string frame is given by
\begin{equation}
\label{line}
ds^2 =-dt^2 +h_{ab}\omega^a \omega^b  ,
\end{equation}
where $\omega^a$ are the one--forms for 
each Bianchi type. The three--metric may be 
diagonalized: 
\begin{equation}
\label{diagonal}
h_{ab}(t) =e^{2\alpha(t)} \left( e^{2\beta (t)} 
\right)_{ab}  ,
\end{equation}
where 
\begin{equation}
\label{beta}
\beta_{ab} ={\rm diag} \left[ -2\beta_+ , \beta_+ -\sqrt{3} 
\beta_- , \beta_+ + \sqrt{3} \beta_- \right]
\end{equation}
is a traceless matrix that determines the anisotropy in the models and
$e^\alpha$ is the `averaged' scale factor. Under the conformal
transformation given in Eq.~(\ref{conformal}), the averaged scale
factor in the Einstein frame is given by
$\tilde\alpha=\alpha-\varphi/2$, but the definition of the variables
$\beta_\pm$ is conformally invariant.  

The antisymmetric structure constants, 
${C^a}_{bc}$, of the Lie algebra of $G_3$ may be written as 
${C^a}_{bc}=
m^{ad}\epsilon_{dbc} +{\delta^a}_{[b}a_{c]}$, where
$m^{ab}=m^{ba}$, $a_c \equiv {C^a}_{ac}$,
$\epsilon_{abc}$ is a covariantly 
constant three--form 
and indices are raised and 
lowered with $h^{ab}$ and $h_{ab}$, respectively.
The Jacobi identity then 
implies that $m^{ab}a_b =0$ and the Lie algebra is in the 
Bianchi class A or B if $a_b=0$ or $a_b \ne 0$, 
respectively \cite{EM69}. The Lagrangian formulation 
of the class B models 
is ambiguous because the condition $a_b \ne 0$ 
leads to non--vanishing spatial divergence terms. 
Such problems do not arise in the class A, however, 
and this includes the Bianchi types 
I, II, ${\rm VI}_0$, ${\rm VII}_0$, 
VIII and IX. We therefore restrict our discussion 
to these models\footnote{It should be remarked, however, 
that the analysis is also valid for 
Bianchi types III and V 
and the Kantowski--Sachs model. See, e.g., Ref.~\cite{WE97} for 
a full discussion of the Bianchi types.}. 

A Lagrangian for the Bianchi class A models may be derived from
the dual action (\ref{solitonicaction}) by integrating over the
spatial variables. If the scalar fields are themselves homogeneous on
the spatial hypersurfaces, it follows that the action can be written as
\begin{equation}
S_{4*} = \int dT \, \left[ L_g + L_m \right] \ , 
\end{equation}
where 
\begin{eqnarray}
\label{gravaction}
L_g &\equiv& -6\dot{\tilde{\alpha}}^2 
+ 6 \dot{\beta}_+^2 +6\dot{\beta}_-^2+ {^{(3)}}R (\tilde{\alpha}, 
\beta_{\pm} ) \ , \\
\label{matteraction}
L_m &\equiv& \frac{1}{2} \dot{\varphi}^2 
+\frac{1}{2} \dot{y}^2 +\frac{1}{2}
e^{\sqrt{3} y+\varphi} \dot{\chi}^2 
+\frac{1}{2} e^{-\sqrt{3} y +
\varphi} \dot{\sigma}_2^2 +
\frac{1}{2} e^{2\varphi} \left( \dot{\sigma}_1
-\chi \dot{\sigma}_2 \right)^2
\ ,
\end{eqnarray}
represent the gravitational and matter Lagrangians, 
respectively, a dot denotes differentiation with 
respect to the time coordinate in the Taub gauge~\cite{Taub51}
\begin{equation}
\label{taubtime}
T\equiv \int d\tilde{t} \, e^{-3\tilde{\alpha}}
 = \int dt \, e^{-3\alpha-\varphi} \ ,
\end{equation}
and the scalar curvature of the three--surfaces is 
given by ${^{(3)}} R =-(m_{ab}m^{ab} -(m^a_a)^2/2)$.

It follows, therefore, that the action for each specific Bianchi type 
is uniquely determined once the functional form 
of the three--curvature has been specified. 
The advantage of employing the time variable (\ref{taubtime}) 
is that the gravitational and matter sectors of the action 
(\ref{solitonicaction}) are effectively decoupled from one another. 
This implies that 
the vacuum solutions $( \dot\chi = \dot\sigma_i =0)$ for 
the dilaton and moduli fields are given linearly by 
\begin{eqnarray}
\label{dilatonvacuum2}
\varphi &=& \varphi_* + \left( \sqrt{2E} \cos\xi_* \right)  \, T \ ,\\
\label{dilatonvacuum3}
y &=& y_* + \left( \sqrt{2E} \sin\xi_* \right) \, T \ ,
\end{eqnarray}
for all Bianchi types, where $E$ is an arbitrary positive constant of
integration which represents the kinetic energy associated with
Lagrangian $L_m$. Consequently, these solutions  
correspond to straight-line trajectories in the 
($\varphi$,$y$) field space. They are 
tilted at an arbitrary angle $\xi_*$ to the
$\varphi$--axis.  
Note that the averaged scale factor in the Einstein frame,
$e^{\tilde{\alpha}}$, is the solution for a homogeneous cosmology
with a stiff perfect fluid equation 
of state~\cite{MW95}. For instance,  in the case of 
the spatially flat, Bianchi type I model,  we have
$\tilde\alpha\propto T$, 
and the scale factor in the string frame is then given by
$e^\alpha = e^{\tilde{\alpha}+(\varphi/2)}$.

\subsection{Single form field}

An SL(2,R) transformation given by Eq.~(\ref{sl2rchi})
or~(\ref{sl2rsigma2}) generates the general solution with a single,
non--trivial form field $\chi$ or $\sigma_2$, respectively, when
applied to the general homogeneous dilaton--moduli--vacuum
solution~\cite{CLW94,prep,luovwa,lmp,CLW97,kaloper}. If initially $\chi
=\sigma_1=\sigma_2 =0$, then the SL(2,R)$_\chi$ transformation
(\ref{sl2rchi}) generates a single time--like, covariantly conserved
four--current $L_\mu$, and therefore a conserved charge
$L=e^{\varphi+\sqrt{3}y}\dot\chi=2|CD|\cos(\xi_*-\pi/3)$.  This places
a lower limit on the variable $u$, defined in Eq.~(\ref{ufield}), and
thus the 10-D dilaton, $\Phi$.  On the other hand, the
SL(2,R)$_{\sigma_2}$ transformation~(\ref{sl2rsigma2}) generates a
conserved charge
$J=e^{\varphi-\sqrt{3}y}\dot\sigma_2=2|C'D'|\cos(\xi_*+\pi/3)$,
placing a lower limit on the variable $w$ defined in
Eq.~(\ref{wfield}).

Since the matter component (\ref{matteraction}) of the dual action has
the same form for all Bianchi types, we may conclude that
the trajectories of the corresponding solutions in the $(\varphi ,
y)$ plane are identical to those for the isotropic FRW solutions
\cite{prep}. All solutions with a single excited form field exhibit a
single bounce between asymptotic dilaton--moduli--vacuum
solutions. The coupling of the form field to the dilaton and moduli
determines the reflection angle of the bounce~\cite{prep}.

\subsection{Two form fields}

The field equations imply that the only consistent solution with one 
trivial and two non--trivial form fields corresponds to the case where 
the field strength of the NS--NS field vanishes, 
$H^{(1)}=0$. Thus, the pseudo--scalar axion 
field $\sigma_1$ may be eliminated by employing
the definition (\ref{defH1}), which implies that 
$\dot{\sigma}_1 =\chi \dot{\sigma_2}$. It then follows that the 
two non--trivial form fields become decoupled. Thus, 
their field equations may be directly integrated 
to yield $e^{\varphi+\sqrt{3}y}\dot\chi=L$ and
$e^{\varphi-\sqrt{3}y}\dot\sigma_2=J$, as in the case of only a
single non-trivial form field described above.

The field equations for the dilaton and moduli fields 
may be written in the form of the SU(3) Toda system~\cite{toda}. 
Defining the new variables
\begin{equation}
q_{\pm} \equiv \varphi \pm {y \over \sqrt{3}}
\end{equation}
the matter Lagrangian given in Eq.~(\ref{matteraction}) becomes
\begin{equation}
\label{TodaL}
L_m = 
{1\over8} \left( \dot{q}_+ + \dot{q}_- \right)^2
+ {3\over8} \left( \dot{q}_+ - \dot{q}_- \right)^2 - V(q_\pm) \ ,
\end{equation}
where 
\begin{equation}
\label{Vpot}
V(q_\pm) \equiv {1\over2}\left(J^2 e^{q_+ -2q_-} + L^2 e^{q_- -2q_+}\right)
\ 
\end{equation}
represents an effective potential. 
This implies that the field equations take the form 
\begin{eqnarray}
\label{fieldq1}
\ddot{q}_- =J^2 e^{q_+ -2q_-} \\
\label{fieldq2}
\ddot{q}_+ = L^2 e^{q_- -2q_+}
\end{eqnarray}
subject to the constraint equation 
\begin{equation}
\label{friedmannq}
{1\over8} \left( \dot{q}_+ + \dot{q}_- \right)^2
+ {3\over8} \left( \dot{q}_+ - \dot{q}_- \right)^2 + V
= E,
\end{equation}
where $E$ is a positive constant.

The general solution to Eqs. (\ref{fieldq1})--(\ref{friedmannq})
is of the form~\cite{kaloper,luovwa,lmp,prep}
\begin{equation}
e^{q_-} = \sum_{i=1}^3 A_i e^{-\lambda_iT} \ , \quad
e^{q_+} = \sum_{i=1}^3 B_i e^{\lambda_iT} \ ,
\end{equation}
where $\sum_i\lambda_i=0$. This implies that $\lambda_{\rm
min}<0$ and $\lambda_{\rm max}>0$ and the 
asymptotic solution for $\varphi$ and $y$ as
$T\to-\infty$ is therefore given by 
\begin{equation}
\label{a1}
e^\varphi \sim e^{-(\lambda_{\rm max}-\lambda_{\rm min})T/2}
\ ,\quad 
e^y \sim e^{\sqrt{3}(\lambda_{\rm max}+\lambda_{\rm min})T/2}  .
\end{equation}
On the other hand, we have  
\begin{equation}
\label{a2}
e^\varphi \sim e^{(\lambda_{\rm max}-\lambda_{\rm min})T/2}
\ ,\quad
e^y \sim e^{\sqrt{3}(\lambda_{\rm max}+\lambda_{\rm min})T/2}  
\end{equation}
in the limit $T\to+\infty$. 

The asymptotic
solutions (\ref{a1}) and (\ref{a2}) 
correspond to straight lines in the $(\varphi,y)$ plane. 
This is similar behaviour to that of the 
single form field solutions discussed above. More 
specifically, 
the incoming trajectories arrive from infinity ($\varphi\to\infty$) 
at an angle $\xi_*$ to the $\varphi$--axis and 
are reflected back at an angle $-\xi_*$.
The constraint equation (\ref{friedmannq}) implies 
the upper bound,  $V\leq E$, on the potential  
(\ref{Vpot}) and this restricts the range of 
$\xi_*$ to $| \xi_* | \le \pi /6$ \cite{prep}.

\subsection{Three form fields}

We wish to investigate here the general cosmological solution with all
three form fields non-trivial \cite{prep}.  
The general FRW solutions to the type
IIB string action presented in Eq.~(\ref{solitonicaction}) can be
obtained from the dilaton--vacuum solutions by a general SL(3,R) 
transformation. This is equivalent to three (non--commuting)
SL(2,R) transformations:
SL(2,R)$_\chi\times$SL(2,R)$_{\sigma_2}\times$SL(2,R)$_\chi$ which 
yields a complicated analytic form.  
The general solution exhibits a sequence of bounces between asymptotic
vacuum states. We find that the time--dependence of the fields $\chi$ and
$\sigma_2$ induces lower bounds on the variables $u$ and $w$,
respectively, as seen in the single form--field solutions.  
In the general solution this results in a
lower bound on $\varphi = u+w$.

The general type IIB solution contains a non-vanishing NS--NS
form--field, but can always be obtained from the Toda system of
Eq.~(\ref{TodaL}) where $H^{(1)}=0$ by a single SL(2,R) transformation
(\ref{sl2rchi}).  The asymptotic behaviour of $\varphi$ and $y$ is
invariant under this transformation. This follows since $u\to\infty$
asymptotically for all solutions in the Toda system\footnote{An
exceptional case is when $u\to u_*$ asymptotically, where $u_*$ is a
constant. In this case $\bar{u}\to$constant, though not necessarily
$u_*$, but the qualitative behaviour is the same.}  and, from
Eq.~(\ref{newu}), we obtain $\bar{u}\to u$ in the general solution. We
also have $\bar{v}=v$ and thus $\varphi$ and $y$ are invariant in this
limit. Thus, trajectories in $(\varphi,y)$ field-space come in at an
angle $\xi_*$ and leave at an angle $-\xi_*$, where
$-\pi/6\leq\xi_*\leq\pi/6$.

\section{Discussion}

In this paper we have considered a 4--D effective action derived by
compactifying the 10--D type IIB supergravity action 
on an isotropic six--torus. Only the variations
of the form fields on the external 4--D spacetime were considered
and, although many
degrees of freedom present in the full 10--D theory were omitted, 
the interactions between the form fields 
of the NS--NS and RR sectors were included. 
These interactions are such that a dual
action, Eq. (\ref{solitonicaction}),
may be derived, where the field strengths of the two--forms are
interchanged with the one--form field strengths of two pseudo--scalar
axion fields. The five scalar fields in this action 
parametrize an SL(3,R)/SO(3) coset. 

The global symmetry of the 10--D effective action, ${\rm
SL(2,R)}_{\chi}$, is respected by the dual action. However, this
action is also invariant under a discrete ${\rm Z}_2$ reflection
symmetry and this implies the existence of a second, global SL(2,R)
symmetry, ${\rm SL(2,R)}_{\sigma_2}$, corresponding to the mirror
image of the original SL(2,R)$_\chi$ symmetry.
This results in an effective interaction potential for the dilaton,
modulus and RR axion fields that may be expressed in a SL(2,R)
invariant form.
The ${\rm SL(2,R)}_{\chi}$ symmetry mixes the RR axion and 10--D
dilaton and leaves the internal space in the 10--D Einstein frame
invariant. Thus, both the 10--D and 4--D Einstein frame metrics
transform as singlets under this symmetry.  However, the Z$_2$
symmetry is a `T-duality' which inverts the size of the internal
space.  Although the 4--D Einstein metric remains invariant, the 10--D
Einstein metric is {\em not} invariant under the Z$_2$ (or the
resulting SL(2,R)$_{\sigma_2}$) symmetry.

It is interesting to compare the symmetries of the dual action 
(\ref{solitonicaction}) with the more familiar symmetries of the
NS--NS sector of the string effective action.
It is well known that 
there is an SL(2,R) symmetry between the dilaton and the NS--NS
axion when the RR form--fields are zero~\cite{NSdual}. The NS--NS
three--form $H^{(1)}_{\mu\nu\lambda}$ is dual to
$e^{2\varphi}\nabla_{\kappa}\sigma_1$ in this case  and
Eq.~(\ref{solitonicaction}) can be written as
\begin{equation}
\label{NSNSaction}
S_{4*} = \int d^4 x \sqrt{-g} \left[ R -\frac{1}{2}  \left( 
\nabla y \right)^2 + \frac{1}{4} {\rm 
Tr} \left( \nabla N \nabla N^{-1} \right) 
\right] ,
\end{equation}
where
\begin{equation}
N \equiv    \left( \begin{array}{cc} e^{\varphi} 
&  \sigma_1 
e^{\varphi} \\ 
\sigma_1  e^{\varphi} & e^{-\varphi} +
\sigma_1^2 e^{\varphi} \end{array} \right)
\end{equation}
is a member of the group ${\rm SL(2,R)}$. 
Thus,  the NS--NS sector is invariant under a global SL(2,R)
transformation~\cite{NSdual}
\begin{equation}
\label{sl2rsigma1}
\bar{N}=\Sigma N\Sigma^T , \qquad
\bar{g}_{\mu\nu}=g_{\mu\nu} , \qquad
\bar{y}=y 
\end{equation}
and there is also an independent Z$_2$ symmetry:
\begin{equation}
\bar{y} = -y   .
\end{equation}
The SL(2,R) symmetry (\ref{sl2rsigma1}) 
is not respected by the terms involving the RR field
strengths, but there is a striking similarity between the symmetry of
the complete dual action~(\ref{solitonicaction})
and the restricted NS--NS action~(\ref{NSNSaction}). 
In the case of the NS--NS sector alone, the SL(2,R) and Z$_2$
symmetries act on different fields and thus commute, whereas this is
not so in the type IIB effective action (\ref{solitonicaction}). In
the former case, the SL(2,R) symmetry is its own mirror image under
the Z$_2$ reflection, but when the RR fields are included, we find
that the mirror image of the original SL(2,R)$_\chi$ symmetry is a
second SL(2,R)$_{\sigma_2}$ symmetry.
Both actions include an SL(2,Z) S--duality and a discrete Z$_2$
T--duality, and we have seen that many qualitative features of the
NS--NS solutions alone~\cite{CLW94} are shared by the more general type IIB
solutions with both NS--NS and RR fields included~\cite{prep}.

The action (\ref{solitonicaction}) corresponds to a truncation 
of the bosonic sector of $N=8$ supergravity. Although the U--duality 
group of this theory is known to be the full ${\rm E}_{7(7)}$ 
group, the advantage of employing the truncated action is that 
exact solutions to the field equations may be derived. 
We have demonstrated in 
Section~IV that these symmetries of the dual action are powerful tools
for generating new 4--D backgrounds with non--trivial RR fields from
vacuum solutions or solutions where only the NS--NS sector
is non--trivial. 
The cosmological solutions at early or late times approach the
dilaton--moduli--vacuum solutions asymptotically where the NS--NS and RR
form fields become fixed. However the presence of the form--fields
restricts the allowed range of the asymptotic solutions. It is
interesting to note that the pre-Big Bang scenario proposed by
Gasperini and Veneziano~\cite{pBB} is based on solutions to the
low-energy string effective action that emerge from the weakly--coupled
dilaton--moduli--vacuum. These solutions are not typical of the general
axion--dilaton--moduli solutions where the form fields place a lower
bound on the dilaton coupling. Even if the form fields are set to zero
then we expect quantum fluctuations with spectra that can be computed
assuming the Bunch-Davies vacuum state in the short-wavelength
limit. The SL(2,R) symmetry of the NS--NS sector previously has been
used to obtain the perturbation spectra of the axion and dilaton
fields \cite{CEW97,CLW97} and a similar technique using the SL(3,R)
symmetry could be applied to include perturbations in the RR sector.

More generally, the symmetries discussed in this paper may be employed
in conjuction with previous analyses of non--compact symmetries to
generate a wide class of anisotropic and inhomogeneous cosmologies or
membrane solutions from a given vacuum or NS--NS solution
\cite{PS,Feinstein,anisotropic,p,inhomogeneous}.  A solution
containing excited dilaton and axion fields may be found by applying
an ${\rm O}(d,d)$ transformation to an appropriate vacuum solution of
Einstein gravity.  This is relevant to the study of inhomogeneous
string cosmologies where the homogeneity is broken along one
direction. Such models are known as Einstein--Rosen solutions and
admit two abelian isometries and, therefore, a global ${\rm O}(2,2)$
symmetry \cite{ver,bakas}.  Beginning with a vacuum Einstein--Rosen
solution, an ${\rm O}(2,2)$ transformation, followed by the SL(2,R)
transformation (\ref{sl2rsigma1}), would lead to a general class of
inhomogeneous NS--NS cosmologies. These solutions would then serve as
the starting point for generating the RR sector via the symmetries
discussed in Section III. A similar procedure may be employed to
generate plane wave solutions involving RR fields \cite{kar}.
Solutions of this nature are interesting because they model the
collision of gravitational waves in the very early
universe~\cite{grif}.  They would therefore allow the effects of
stringy excitations on the propagation of tensor perturbations to be
investigated.

\vspace{.3in}
\section*{Acknowledgements}
EJC and JEL are supported by the Particle Physics and Astronomy Research
Council (PPARC), UK.  We thank J. Maharana for useful discussions.

\end{document}